\documentclass{article}

\PassOptionsToPackage{numbers, compress}{natbib}

\usepackage[final]{neurips_2020}

\usepackage[utf8]{inputenc} 
\usepackage[T1]{fontenc}    
\usepackage{hyperref}       
\usepackage{url}            
\usepackage{booktabs}       
\usepackage{amsfonts}       
\usepackage{nicefrac}       
\usepackage{microtype}      
\usepackage{graphicx}

\title{Training Ethically Responsible AI Researchers: a Case Study}

\author{%

  Hang Yuan\thanks{Equal contributions.} \textsuperscript{1, 2}, 
  Claudia Vanea\footnotemark[1] \textsuperscript{2, 3},
  Federica Lucivero\textsuperscript{1, 4}, 
  Nina Hallowell\textsuperscript{1, 4}\\
  Nuffield Department of Population Health\textsuperscript{1}\\
  Department of Computer Science \textsuperscript{2}\\
  Nuffield Department of Women's Reproductive Health\textsuperscript{3}\\
  Ethox and Welcome Centre for Ethics and Humanities\textsuperscript{4}\\
  University of Oxford\\
  \texttt{hang.yuan@cs.ox.ac.uk, claudia.vanea@wrh.ox.ac.uk,} \\  \texttt{federica.lucivero@ethox.ox.ac.uk,nina.hallowell@ethox.ox.ac.uk} 
}

\begin{document}

\maketitle

\begin{abstract}
Ethical oversight of AI research is beset by a number of problems. There are numerous ways to tackle these problems, however, they leave full responsibility for ethical reflection in the hands of review boards and committees. In this paper, we propose an alternative solution: the training of ethically responsible AI researchers. We showcase this solution through a case study of a centre for doctoral training and outline how ethics training is structured in the program. We go on to present two second-year students’ reflections on their training which demonstrates some of their newly found capabilities as ethically responsible researchers.
\end{abstract}

\section{Introduction}

As noted in a recent paper by \citet{friesengoverning} ethical oversight of AI research is beset by a number of problems. The first problem is that Institutional Review Boards and Research Ethics Committees (IRBS and RECS) who review biomedical (and social) research involving AI, often lack the requisite technical expertise. The second is that internal (i.e. industry-funded) ethics committees who oversee research undertaken in the commercial sector may have a conflict of interest. There are a number of ways in which we might solve these problems by a) recruiting data scientists to IRBs/RECs, b) providing specialist training for committee members and c) by ensuring greater transparency and more accountable decision-making in industry ethics committees. In this paper, we argue that we could adopt an alternative solution, one that does not put the responsibility for ethics purely in the hands of ethical oversight committees, but requires that all research involving AI is ethical by design. In other words, we suggest that all those who work in this domain should have greater ethical awareness and that this may be achieved by training researchers in ethics.

Producing ethically aware and responsible health data scientists to work on AI in health is the goal of the EPSRC-funded Centre for Doctoral Training (CDT) in Health Data Science at the University of Oxford. The CDT aims to produce data scientists who are not only aware that their research may have an ethical impact on individuals and society, but also are trained to design ethically robust responsible research projects. In this paper, we describe how ethics is embedded within the training that students receive and how this training has affected the ways in which some of our students now approach their research.

\section{Ethics in the CDT}

The CDT’s curriculum includes ethics training at different points over the four-year course. Despite the students' differing backgrounds (mainly mathematics, statistics, and computer science), when designing the course, we were clear that we did not want them to just identify some of the ethical issues arising in AI research, but also wanted to ensure that they learnt how to apply their training and assess the ethical issues in  their own and  others’ research. To encourage students to embrace ethics, we have embedded ethics at different points in the course. Students have three lecture courses about: a) \textbf{Information/data governance} in which they learn about the legal rules around the data protection, b) \textbf{Contemporary topics in the ethics of data science} (e.g. ethics of data sharing, social justice, responsible research, algorithmic bias, environmental impact of data science, misuse, etc.) and c) \textbf{Responsible Research \& Innovation (RR\&I)}. These are designed to increase understanding about data ethics and data protection, encourage critical thinking and, in the case of the RR\&I course, provide them with a framework to test the ethical robustness of their research ideas. 

Students are then required to apply their ethical learning and the tools we have provided to the projects they undertake within the CDT. In the first year they do two group-based data challenges and two individual projects, and, in every case, have to present an assessment of the ethical impact of the project. They are expected to consider the harms and benefits, how the former may be mitigated and the latter increased, and consider the broader impact on individuals and society. When designing their doctoral research, they are required to think about the ethical issues it may raise and present these as an ethics impact assessment to both their supervisors and the class halfway through their second year. We hope by encouraging them to consider these issues and apply them to their own research, the students will not only see their research through an ethical lens, but also will create research projects that involve ethical thinking from the beginning. Ultimately, the goal of the ethics stream in the CDT is to make students ask not only what I can do in my research, but also what should I do? 

This shift in thinking was partially achieved through class discussion of ethics case studies and reflections on their ongoing research practice. For example, in one of the lectures, “\textbf{Contemporary topics in the ethics of data science}”, students were introduced to the ethical debate on the use of digital tools in healthcare. They were asked to reflect on the data challenge they had been previously involved in and to engage in a debate focusing on two sets of questions: a) How the exercise challenged some of the assumptions they had about doing research and how, from their perspective, this data rich research changes the way of doing science and developing knowledge? b) What type of ethical considerations they had encountered during the data challenge? 

By asking them to draw on their recent research experience, students could participate by focusing on concrete examples. Moreover, students were presented with two case studies that were a readaptation of challenges that had emerged in real-life research projects. In both cases, they were encouraged to imagine themselves as the researchers in these projects and explore the ethical challenges from that perspective. The goal of the exercise was to trigger their “moral imagination” \citep{fesmire2003john} and encourage them to articulate moral problems that they would expect to encounter. Instead of agreeing on a solution for these moral problems, students were encouraged to dissect the structure of the ethical problems, highlight their assumptions and describe their weakness. This class, and others in the course, were aimed at contextualising students’ ongoing research projects and pushed them to explore ethical issues arising in their research by learning how to provide and analyse ethical arguments.

In sum, students were taught how to approach the complexity of ethical problems and were introduced to a wide variety of likely research scenarios to prepare them for their future work. In the following sections, two second-year CDT students will reflect on the first year of their ethics training. They will present their main takeaways from this part of the program, how these apply to the research pipeline, and their new understanding of the ethical landscape. In addition, they will highlight what they perceived as a common limitation to solving ethical problems and propose a solution.

\section{Student reflection 1: Exploring ethics in the research pipeline}

This training has enabled us to understand and explore the complex ethics ecosystem \citep{samuel2019ethics} of AI research in health. We have learnt, on the one hand, to identify ethical issues along the pipeline of AI research: namely, \textbf{research design, development and deployment}, and on the other hand, to reflect on the potential ethical impact of our work on the people and institutions involved.

For example, there are many cases where the research objectives themselves are questionable \citep{hill2019genome, wang2018deep}. These issues can be dealt with during the \textbf{research design} phase, by asking what might be the outcomes of this research, and to what uses could it be put? In the \textbf{research development} phase, we face a whole different set of ethical questions. One example could be how we can minimise the use of computing power to train our models and reduce the environmental impact of our research, and how we should interpret the modelling outcomes so that we do not bias our conclusions. Finally, when it comes to \textbf{research deployment}, there are other ethical aspects to consider. Let us use GitHub as an example, a tool many use for collaboration and software version control. Using GitHub, one could inadvertently include login credentials in a publicly accessible codebase, rendering research participants’ data available to anyone with an internet connection \citep{ursemdata2020}. Hence, even for a choice as trivial as a version control tool, we need to think about the potential ethical risks and whether these risks are justified/could be attenuated given the benefits it provides. The above list of ethical problems is not meant to be exhaustive but showcases the types of ethical decisions researchers need to make at different points in the research process. Arguably, it is only by identifying these issues that we can think about the best ways to resolve them and therefore avoid the potential harms.

Working in the medical AI space, we will naturally impact individuals at different stages of the research pipeline. For example, during the research design phase, we need to outline  and implement ethical consent practices when personally identifiable data is needed, or during the development and deployment phase, we need to think about the suitability of reporting incidental disease findings. Indeed, throughout different stages of research, we need to be cautious about the use of individuals' data and endeavour to ensure that we do not exploit the trust that participants have put in us. 

In addition to the research participants, our research may also have a profound impact on society. Let us use \citet{hill2019genome} as an example, where a claimed link between individuals' genetics and socioeconomic status could become a self-fulling prophecy, stigmatising groups of individuals. Another example is the Cambridge Analytica case in which Facebook data was used to target voters and influence voter attitudes in the US presidential elections. It is likely that the individuals who developed the technologies used in these cases did not anticipate that their (seemingly innocuous) work would be used with malicious intent. Even if we could predict all possible detrimental influences our work might have and could devise means to circumvent them, there is no guarantee that our solutions would be bullet-proof. However, we would argue that considering the potential misuse of research and putting mitigation in place is a small step towards making our research more ethical. 

The examples above demonstrate the range and types of ethical issues and questions that all researchers need to consider. Of course, individual researchers are not the only actors and responsibilities also lie with institutional bodies (universities, funding bodies and laboratories) whose decisions and policies raise ethical questions. As a consequence of this, when conducting ethical research, we may find ourselves unable to please everyone all the time. This is a challenging reality that researchers need to navigate even when they intend to do the ``right thing”.
\section{Student reflection 2: Dealing with ethical uncertainty in research}

Knowing what the ethical problems/questions are is one thing, knowing how to solve them is another. As AI researchers, we are used to searching for the most optimal solution to a particular problem. Through experimentation, we narrow down our search space until we are measurably closer to a solution. Our models’ accuracy improve, our losses decrease, we gain computational efficiency, and so on. There is a path which conclusively leads to our desired result; we accept or reject our pre-established null hypotheses. However, as was soon apparent from the many inconclusive ethical debates sparked during our training, finding optimal solutions to ethical questions is not straightforward. There are no statistically verifiable, pre-established hypotheses and ethics cannot be so easily measured \citep{bentham1996collected}. Nonetheless, this does not mean that ethical problems have no solution and a major takeaway from our training was how we might attempt to construct a solution.

The first step towards constructing a solution is, unsurprisingly, gaining a better understanding of the problem. For ethical questions, this will involve having an awareness of the broader implications of our actions, the severity of those implications, any precedents relating to those actions, rules and regulations, and so on. Sometimes, these factors will lead to a clear solution; for example, if we wish to use personally identifiable data, then there are rules to follow for obtaining the necessary informed consent. Other times, these factors will introduce uncertainty or be contradictory, however, they can be used to highlight the limitations of proposed solutions. By understanding the ethical landscape surrounding the question, we can gain some clarity with which to evaluate proposed solutions. Additionally, and perhaps most importantly, if we are unsure about the morality of a decision at any stage of our research, there are people and/or bodies that we can turn to for advice. 

We may better understand this relationship between uncertainty and clarity by looking at specific examples. Intended more as a tool for thinking than a formal framework, Figure 1 (see Appendix) shows how ethical problems sit at different points along a scale of clarity. For some problems, there are clearly established rules/guidelines, such as obtaining informed consent for the use of personally identifiable data. For other problems, where there is no single best solution, we can at least be certain about some of the things we should not do. For example, it is unclear whether we should disclose to participants any incidental research findings which may affect their health. However, it is clear that a disclosure policy should be decided upon in advance and the participants should be informed of this policy when giving consent. In these examples there is a degree of clarity about what we should do.

Moving down the scale of ethical clarity, we encounter problems where all realistic solutions have considerable trade-offs and there are fewer precedents to follow. As a result, discussions surrounding these kinds of problems often end inconclusively. Take the controversial Stanford sexual orientation study as an example \citep{wang2018deep}. In this project, researchers trained neural networks to identify sexual orientation from facial images scraped from publicly available dating websites. The authors point out that their findings support the idea that sexual orientation may be an innate trait and warn that this kind of technology could be used to identify groups of people with traits that certain societies regard as socially undesirable. However, at the same time, the authors fail to acknowledge that their training data were obtained without individuals’ consent, they fail to account or adjust for inherent biases in the data, and do not consider that the attention generated by their study may have accelerated uptake of the very methodology they warn against. As this example demonstrates, research can generate a number of ethical issues, so an exact solution is less clear.

Finally, even a seemingly ethically neutral problem where one might expect total freedom of choice, such as choosing a project’s programming language, generates ethical considerations. These may include considerations of the suitability of the language to the task, your personal efficiency in the language, the inherent computational burden of the language, and wider community accessibility. Although this example may have less of an ethical impact than some of the problems outlined above, it demonstrates how ethical clarity does not scale with consequence or severity but still requires analysis and understanding of the problem to arrive at the best solution.

In many of these cases, uncertainty when dealing with ethical problems in research can be mitigated, to some extent, by gaining a better understanding of the ethical landscape surrounding the problem and the costs and benefits of possible solutions. For especially complex ethical problems with little ethical clarity and greater impact, we, as students, developed our own solution for dealing with this uncertainty: the moral sense of \emph{being reasonable}. Akin to the approaches proposed by normative theories of virtue or pragmatist ethics \citep{crisp1997virtue, fesmire2003john}, \emph{being reasonable} or \emph{reasonableness}, can be seen as a combination of ethical instinct, informed peer consensus, awareness of cultural norms, and the mitigation of harm coupled with the promotion of benefit. 

Whilst in many ways this approach is still prone to the same issues of subjectivity, inconclusiveness, and uncertainty as was the original ethical problem, it relies on the value judgement of appropriately trained researchers to come to conclusions. Perhaps this reflects the intentions of the CDT to not only train students in the rules and foundations of ethics in health data science, but to instil in them the critical thinking and ethical instinct needed to solve these kinds of problems for themselves.

\section{Conclusion}
The CDT set out to train students in the tools and critical thinking needed to approach ethical problems with the goal of getting students to think about ethics at every stage of their research. Given the student’s reflections, these intentions appear to have been met, however, we will have to wait and see whether it leads to them becoming ethically responsible researchers. Moving beyond the CDT, we encourage, and are happy to see that, more programs are beginning to incorporate ethics into their training schemes. This will allow for future research to empirically evaluate and compare the success of such approaches, in addition to their broader impacts.

\section*{Broader Impact}

In this paper, we have proposed the training of ethically responsible AI researchers as a fundamental step towards more ethical AI research. Instilling in new researchers the importance of being aware of, and providing the tools to understand, ethical problems in AI research will result in a more ethically aware AI research community. This community, in turn, will produce AI research that is ethical at all stages of the research pipeline and will be better equipped to evaluate others’ research as well as their own. A more ethically aware research community should lead to easier ethical oversight processes and their research which is, in theory, ethical by design should reflect a greater awareness of both the benefits and harms of AI based solutions.

There are, however, potential issues with this proposed solution. It relies on good quality, unbiased ethics training, which itself should be subject to review. This may mean that new guidelines, regulations and review boards or committees need to be introduced to evaluate this training, which may reintroduce the original problems found in ethical oversight. Additionally, there are difficulties in ensuring that this training is successful and that all new researchers come away with an appropriate desire to approach research ethically. Finally, of course, the addition of this ethics training to existing research training programs or the introduction of new ethics training for existing researchers will incur substantial costs, be that financial, time, or otherwise.

As more training programs begin to incorporate ethics into their syllabi, time will tell whether this approach really does lead to a more ethically aware research community.

\begin{ack}
Hang Yuan and Claudia Vanea are second-year students in the Health Data Science CDT at Oxford University UK. 
Hang Yuan’s doctoral research is funded by the Li Ka Shing Foundation, Claudia Vanea’s doctoral research is  funded by the EPSRC.
Nina Hallowell is one of the Codirectors of the EPSRC Centre for Doctoral Training in Health Data Science at Oxford University; she has overall responsibility for the Ethics training. Federica Lucivero is a co-organiser of the lecture course Contemporary topics in the ethics of data science in the CDT.

Nina Hallowell is funded by the Li Ka Shing Foundation, both Nina Hallowell and Federica Lucivero are members of of the Ethox Centre and the Wellcome Centre for Ethics and Humanities which is supported by funding from the Wellcome Trust (Grant no 203132)  
\end{ack}

\small
\bibliography{aiImpact}

\newcommand{\noop}[1]{}
\begin{thebibliography}{8}
\providecommand{\natexlab}[1]{#1}
\providecommand{\EM}{\em}
\providecommand{\RNtxt}{\relax}
\RNtxt{}

\bibitem[Bentham(1996)J.~Bentham]{bentham1996collected}
{\EM Bentham Jeremy}.
\newblock The collected works of Jeremy Bentham: An introduction to the
  principles of morals and legislation. 1996.

\bibitem[Crisp, Slote(1997)R.~Crisp, M.~A. Slote]{crisp1997virtue}
{\EM Crisp Roger, Slote Michael~A}.
\newblock Virtue ethics. 1997.

\bibitem[Fesmire(2003)S.~Fesmire]{fesmire2003john}
{\EM Fesmire Steven}.
\newblock John Dewey and moral imagination: Pragmatism in ethics. 2003.

\bibitem[Friesen et~al.(2020)P.~Friesen, R.~Douglas~Jones, M.~Marks, R.~Pierce,
  K.~Fletcher, A.~Mishra, J.~Lorimer, C.~Veliz, N.~Hallowell, M.~Graham,
  et~al.]{friesengoverning}
{\EM Friesen P, Douglas~Jones R, Marks M, Pierce R, Fletcher K, Mishra A,
  Lorimer J, Veliz C, Hallowell N, Graham M, others }.
\newblock Governing AI-driven health research: are IRBs up to the task?
  \allowbreak\newblock// Ethics \& Human Research. 2020.

\bibitem[Hill et~al.(2019)W.~D. Hill, N.~M. Davies, S.~J. Ritchie, N.~G. Skene,
  J.~Bryois, S.~Bell, E.~Di~Angelantonio, D.~J. Roberts, S.~Xueyi, G.~Davies,
  et~al.]{hill2019genome}
{\EM Hill W~David, Davies Neil~M, Ritchie Stuart~J, Skene Nathan~G, Bryois
  Julien, Bell Steven, Di~Angelantonio Emanuele, Roberts David~J, Xueyi Shen,
  Davies Gail, others }.
\newblock Genome-wide analysis identifies molecular systems and 149 genetic
  loci associated with income \allowbreak\newblock// Nature communications.
  2019. 10, 1. 1--16.

\bibitem[Samuel et~al.(2019)G.~Samuel, G.~E. Derrick, T.~van
  Leeuwen]{samuel2019ethics}
{\EM Samuel Gabrielle, Derrick Gemma~E, Leeuwen Thed van}.
\newblock The ethics ecosystem: Personal ethics, network governance and
  regulating actors governing the use of social media research data
  \allowbreak\newblock// Minerva. 2019. 57, 3. 317--343.

\bibitem[Ursem, DataBreaches.net(2020)J.~Ursem,
  DataBreaches.net]{ursemdata2020}
{\EM Ursem Jelle, DataBreaches.net }.
\newblock No need to hack when it's leaking. Aug 2020.

\bibitem[Wang, Kosinski(2018)Y.~Wang, M.~Kosinski]{wang2018deep}
{\EM Wang Yilun, Kosinski Michal}.
\newblock Deep neural networks are more accurate than humans at detecting
  sexual orientation from facial images. \allowbreak\newblock// Journal of
  personality and social psychology. 2018. 114, 2. 246.

\end{thebibliography}
\bibliographystyle{rusnat}

\pagebreak

\section{Appendix}
\begin{figure}[h!]
  \centering
  \includegraphics[scale=0.32]{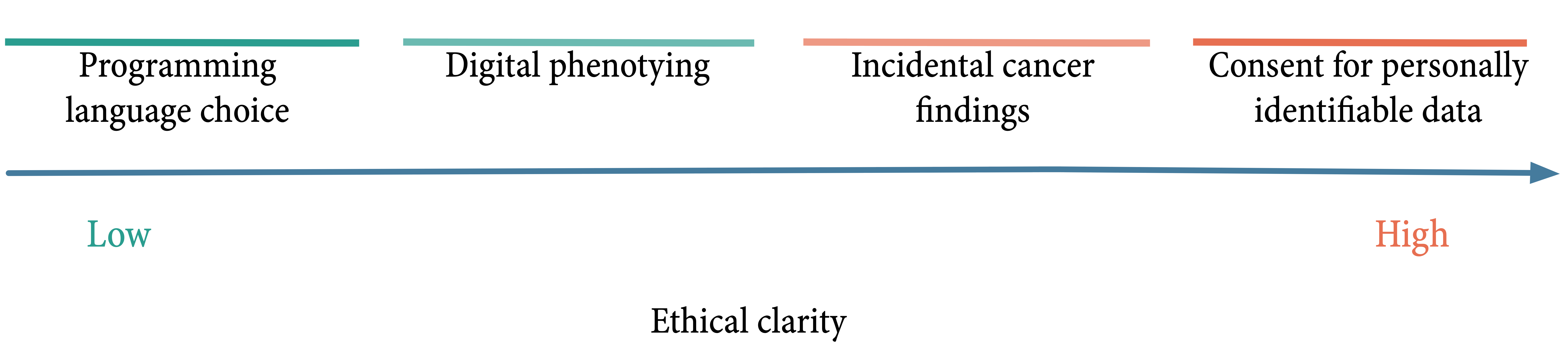}
  \caption{Examples on the scale of ethical clarity.}
\end{figure}
\end{document}